# The Psychological Science of Artificial Intelligence: A Rapidly Emerging Field of Psychology

Zheng Yan[1] Ru-Yuan Zhang[2]
[1]University at Albany [2]Shanghai Jiao Tong University

**Abstract**

The psychological science of artificial intelligence (AI) can be broadly defined as an emerging field of psychology that examines all AI-related mental and behavioral processes from the perspective of psychology. This field has been growing exponentially in the recent decade. This review synthesizes the existing literature on the psychological science of AI with a goal to provide a comprehensive conceptual framework for planning, conducting, and assessing scientific research in the field. It consists of six parts, starting with an overview of the entire field of the psychological science of artificial intelligence, then synthesizing the literature in each of the four specific areas (i.e., Psychology of designing AI, psychology of using AI, AI for examining psychological processes, and AI for advancing psychological methods), and concluding with an outlook on the field in the future.

## 1. Overview: The psychological science of AI

The psychological science of AI can be broadly defined as an emerging field of psychology that examines all AI-related mental and behavioral processes from the perspective of psychology. This field has been growing rapidly in the recent decade. The figure below outlines the field.

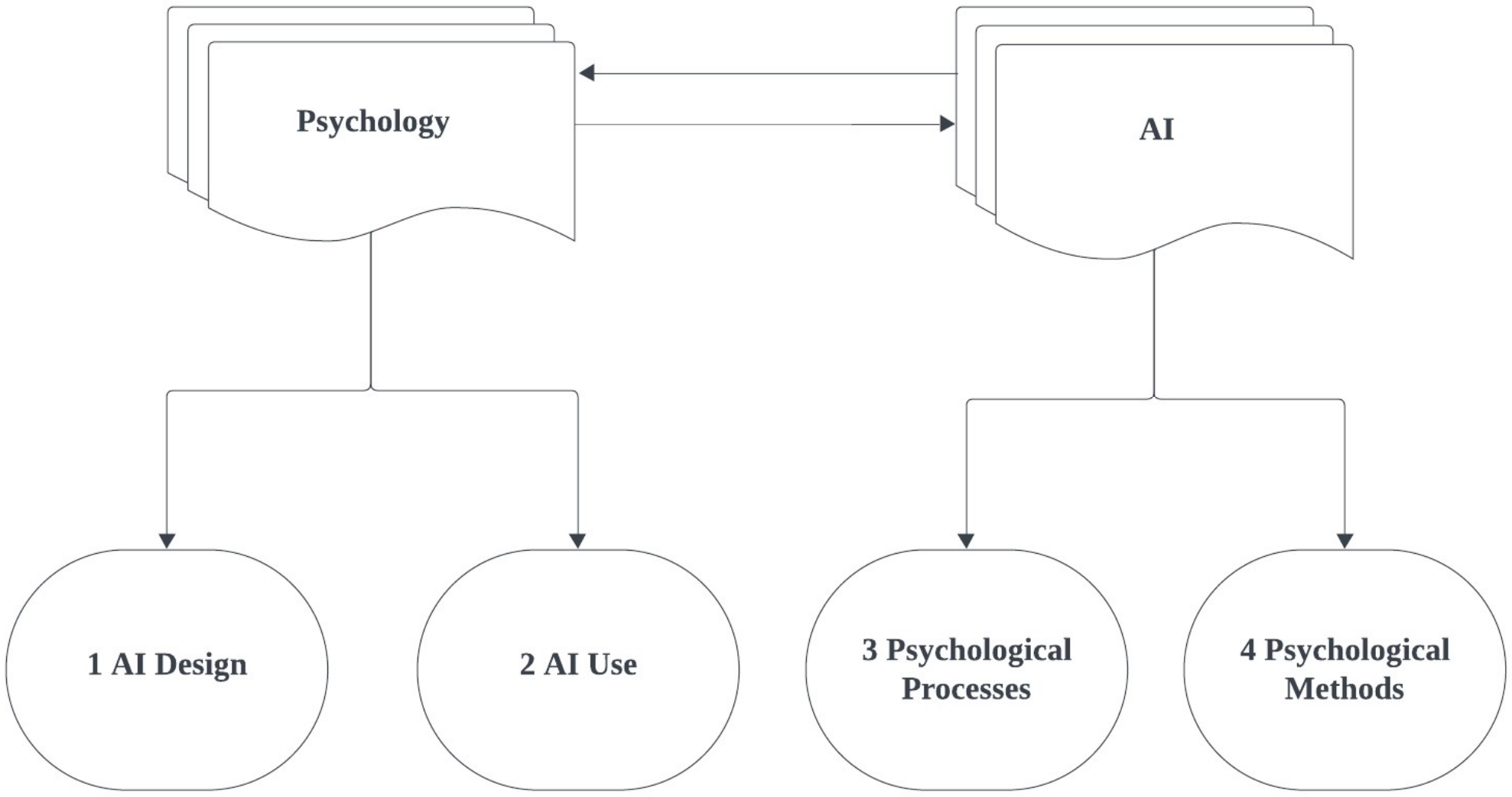

Figure 1. An overview of the psychological science of AI as an emerging field of psychology, with two interconnected systems, Psychology and AI, and four specific areas, Psychology for AI

Design, Psychology for AI use, AI for Psychological Processes, and AI for Psychological Methods.

Figure 1 above indicates five important features of the psychological science of AI as a field of psychology.

First, psychology and AI are **distinctively different** from each other, displayed as two separate systems in Figure 1. Nobel laureate Herbert Simon (1995) called them the psychology side and the AI side. Psychology is often institutionalized as a department in the social sciences, and AI is typically associated with the department of computer science. The existing literature shows that psychology and AI have different definitions, histories, and constituents.

For psychology, it is defined as the scientific study of mind (e.g., perception and thought) and behavior (e.g., walking and talking) of humans and nonhumans by three distinguished professors of psychology at Harvard (Schacter, Gilbert, & Wegner, 2009). It has about 150 years of history since Wilhelm Wundt established the first experimental psychology laboratory in 1879 and has experienced multiple major movements, such as Structuralism, Functionalism, Psychoanalysis, Behaviorism, and the Cognitive Revolution (Schultz & Schultz, 2018). After 150 years of development, **moder psychology** has grown into a particularly large field with various established branches, such as neuropsychology, biological psychology, cognitive psychology, social psychology, developmental psychology, industrial-organizational psychology, educational psychology, and clinical psychology (Myers, 2020). It often includes diverse disciplines that are related to human experiences, such as neuroscience, cognitive science, human-computer interaction, human factors, and behavioral science (Hassabis, Kumaran, Summerfield, & Botvinick, 2017).

Different from psychology, AI was defined by one AI founder John McCarthy (2004) simply as the science and engineering of making intelligent machines. Two eminent AI scientists (Russell & Norvig, 2022) elegantly summarized four features of AI: thinking humanly, acting humanly, thinking rationally, and acting rationally. Unlike psychology's 150 years of history, AI has only 70 years of history since the Dartmouth Summer Workshop in 1956. However, it has already gone through a long roller coaster, with multiple AI Winters and multiple AI Booms, now having the biggest ever AI boom worldwide (Kautz, 2022; Haigh, 2024; Nilsson, 2010). After 70 years of development, **modern AI** now becomes a complex field (Russell & Norvig, 2022). For instance, modern AI highlights a learning-based and data-driven paradigm, which evolved significantly from classic AI featuring a rule-based and logic-driven paradigm; various major algorithms of Modern AI can successfully emulate or even surpass multiple dimensions of human intelligence; worldwide discussions on AI ethics, AI safety, explainable AI, responsible AI, trustworthy AI, and AI governance have been heavily intensified.

Second, although they are distinctively different from each other, psychology and AI are **inherently intertwined** with each other, shown as two-way interactions between psychology and AI in Figure 1. As Nobel laureate Demis Hassabis and his collaborators (2017) stated in their seminal review (Hassabis, Kumaran, Summerfield, & Botvinick, 2017), neuroscience and psychology and AI development have a long-intertwined history. Herbert Simon (1995) also pointed out that AI is a part of computer science and a part of psychology and cognitive science because both inherently deal with intelligence of humans or machines. Specifically, the first basic interaction is "psychology > AI", indicating psychology research can inspire and inform AI technologies. One good example is that the research on reinforcement learning in human brains (Lee, Seo, & Jung, 2012; Ullman, 2019) inspired AI scientists to introduce an algorithm on reinforcement learning to enable AlphaGo to defeat a world champion in the game of Go, even without human data or human guidance (Silver et al., 2017). The second basic interaction is "AI > psychology", showing AI technologies can inspire and inform psychological research. A recent outstanding example is a study by Binz et al. (2025), showing that an AI's foundation model explains human behavior in diverse cognitive tasks and helps build a unified theory of human general cognition. As a result, the psychology of AI emerged essentially at the intersection between psychology and AI.

Third, the basic two-way interactions between psychology and AI can generate **four specific areas**, presented as the four subsystems in Figure 1, indicating that studies in the psychology of AI can be classified into these four areas. **Area 1** is psychology for designing AI, e.g., some humans' cognitive biases (over-relying on the most recent information or making numbers less precise) can inspire AI scientists to design more effective algorithms (Gigerenzer, 2024). **Area 2** is psychology for using AI, e.g., multiple psychological factors such as opacity and rigidity make individuals resist the use of AI (De Freitas, Agarwal, Schmitt, & Haslam, 2023). **Area 3** is AI for understanding psychological processes, e.g., the study of various behaviors of intellectual machines (Does an AI financial product show price discrimination? Does a robot promote good products to children?) can help fully understand animal and human behaviors (Rahwan et al., 2019). **Area 4** is AI for advancing psychological methods, e.g., AI's chatbots can be used to replace human research participants (Hutson, 2023).

There are four important points worth noting here. First, among these four areas, Area 1 (psychology for designing AI) and Area 3 (AI for understanding psychological processes) are **unique** to the psychology of AI. This is because AI essentially is designed to mimic human intelligence and thus certain human-like psychological processes emerge in AI systems. Second, different from Area 1 and Area 3, Area 2 (psychology for using AI) and Area 4 (AI for advancing psychological methods) are **common** in the psychology of computers (Green, Payne, & Veer, 1983), psychology of the Internet (Wallace, 2015), and psychology of mobile phones (Yan, 2015). Third, because of the distinctive difference, not all steps in AI design and AI use are related to psychology, and not all psychological processes and methods are related to AI technologies. However, because of the inherent intertwining between psychology and AI, explicitly or implicitly, directly or indirectly, various steps in AI design and AI use are **ubiquitously** related to psychology, and various psychological processes and methods are ubiquitously related to AI technologies. Fourth, because AI and psychology are inherently intertwined, there often exist **interconnections** across the areas (e.g., Area 1 could be connected with Area 3, and Area 3 could be connected with Area 4). These four areas can be further broken down into **sub-areas** (e.g., Area 3 mainly concerns two types of psychological processes of AI and humans).

Fourth, the entire Figure 1 offers **a conceptual map** of the psychology of AI, depicting the current intellectual landscape of a rapidly emerging field. In recent years, publications on various AI-related psychological phenomena and underlying processes have been surging unprecedentedly. For example, two core journals in the psychology of AI launched two new specialized journals, Computers in Human Behavior: Artificial Humans, in 2023 and Computers and Education: Artificial Intelligence, in 2020. This conceptual map of the psychology of AI can help researchers **understand** relationships of various research efforts (e.g., a study on AI trust and a study on AI acceptance both are related to Area 2 regarding psychology for AI use), **identify** scientific contributions of various studies (e.g., research on AI consciousness can primarily contribute to Area 3 regarding AI for examining psychological processes), and **envision** possible future research programs (e.g., experimental studies in Area 4 regarding AI for advancing psychological methods are much needed, given extensive literature in the area are dominantly conceptual or descriptive). It is timely and useful to have such a conceptual map of the field.

Fifth, Figure 1 also specifies a **specific name,** *the psychological science of AI*, for the field. Over the past 70 years, a substantial body of empirical research has explored the mental and behavioral aspects of artificial intelligence from the perspective of psychology (e.g., Chiarella et al., 2022; Gillath et al., 2021; Isaacet al., 2024; Leichtmann et al., 2023; Pelau, Dabija, & Ene, 2021; Sundar, Russell-Rose, Kruschwitz, & Machleit, 2025). In recent years, scholars have begun to propose the concept of the psychology of artificial intelligence as a field, focusing on Area 3 of using AI to examine psychological process (Friedenberg, 2008; Louwerse, 2025) or on both Area 3 and Area 2 of using psychology to inform AI use (Crowder & Friess, 2012; Prescott, 2024; Williams & Lim, 2024). To distinguish from these existing specific psychologies of AI, we will use the *psychological science of AI* as an umbrella term to label the general psychology of AI that embraces the four specific areas of research.

## 2. Psychology for designing AI (Area 1)

### 2.1 The three phases of the life cycle

According to the two well-established scholars of AI (De Silva & Alahakoon, 2022), designing AI technologies is a complex life cycle and normally consists of three phases through 19 steps. The first phase is **Design** by senior AI scientists. It includes 6 steps: identify and formulate the problem, review data and AI ethics, review technical literature on AI algorithms, prepare data, explore data, and acquire external data. The second phase is **Development** by AI scientists and AI engineers. It consists of 7 steps: pre-process data, build initial AI models, augment data, develop a benchmark, build multiple AI models, evaluate primary metrics, and make AI models explainable. The third phase is **Deployment** by AI engineers. It involves 6 steps: evaluate secondary metrics, deploy AI models with risk assessment, review post deployment, operationalize using AI pipelines, automate processes and systems, and monitor and evaluate performance.

**2.2 The psychology of designing AI**

The entire process of designing AI is essentially an engineering process to build intelligent machines rather than a pure psychology process linked with the various mind and behavior of humans. However, various psychological factors are involved in all three phases of the AI cycle. Psychological research can inspire, inform, and guide AI professionals to design, develop, and deploy AI.

One of the most important articles on psychology for AI design is authored by four neuroscientists and AI scientists (Hassabis, Kumaran, Summerfield, & Botvinick, 2017). In this highly influential review article, Nobel laureate Demis Hassabis and his collaborators have effectively elaborated that psychology, as well as cognitive science and neuroscience, does inspire the design, development, and deployment of AI technologies. Their two central arguments are (1) the fields of neuroscience and artificial intelligence have a long and intertwined history and (2) a better understanding of brains can play a vital role in building intelligent machines. Specifically, they surveyed various historical interactions between AI and neuroscience (e.g., deep learning in neuroscience and AI). They emphasized that neural computation research in attention, episodic memory, working memory, and continual learning has inspired current advances in AI, and highlighted future directions (i.e., efficient learning, transfer learning, imagination and planning, and virtual brain analytics). Clearly, this seminal article focuses on the design of AI algorithms and AI models, critical steps among the 19 steps within the AI design cycle.

The existing literature has documented various studies in the area of the psychology of AI design. Two major contributions of these studies include: (1) **embedded ethics** (Fiske, Henningsen, & Buyx, 2019; McLennanet al., 2020; Morley, Floridi, Kinsey, & Elhalal, 2020; Oudeyer, Kaplan, & Hafner, 2007). For example, McLennanet al., (2020) discussed how to use the psychology of ethics to embed ethics in the entire AI development, especially at the beginning, with two methods, i.e., helping AI developers identify and address various ethical issues and embedding ethicists into the development team. (2) **explainable AI** (Longoet al., 2024; Taylor & Taylor, 2021; Wang, Yang, Abdul, & Lim, 2019). For example, Taylor and Taylor (2021) presented exemplary cases of explainable AI that were inspired by cognitive experimental psychology. They demonstrated that the methods and rigor of cognitive experimental psychology should and could be translated to increase interpretability, fairness, and transparency of AI models.

**2.3 Future research**

Algorithm and model design, ethics design, and explainability design are three research examples of the psychology for AI design. Future efforts are needed to use psychological research to inspire and inform better AI design. Among the three major phases of the AI design cycle, the current research has mainly focused on the phases of design (e.g., algorithm design) and development (e.g., explainability design). Future psychological research should further study the phase of deployment (e.g., assessment design and application design). Future research should study how to understand and train various types of AI professionals (e.g., AI policy makers, AI scientists, AI engineers, AI managers, and even undergraduate and graduate students of AI). For example, to train undergraduates of AI, the ethics of AI was introduced into a computer science course titled "Planning and Learning Methods in AI" in the Paulson School of Engineering and Applied Sciences at Harvard, with the hope to anchor AI ethics into core values of future AI designers."

## 3. Psychology for using AI (Area 2)

Technology for humans generally involves two basic phases sequentially: designing it and using it. Designing AI is to develop AI products with an elite group of AI professionals. In contrast, using AI is to apply AI products in various sectors (e.g., healthcare, business, or education) of the real world, with thousands of AI applied professionals (e.g., technicians working in medical or business schools) and millions of AI public users (e.g., high school teachers and students). Here, the psychology of using AI refers to the study of diverse and complex processes of using AI applications in the real world from the perspective of psychology. Based on the initial search in Web of Science, particularly extensive literature exists, with about 3374 journal articles addressing various psychology-related topics in the process of using AI.

### 3.1 The four basic elements of using AI

Various psychological factors and processes of using a technology involve four basic elements, various technologies, various users, various activities in various domains, and various impacts, based on a basic model of using technologies (Yan, 2012, 2015, 2020, & 2023). In other words, the psychology of using AI should address four basic questions: (1) What specific AI technologies are used? (2) What specific user groups and individuals use AI? (3) What specific activities of using AI are involved in various domains? (4) What specific psychological and behavioral impacts does AI use generate? Like psychological research of using computers, the internet, and mobilephones, psychological research of using AI can inspire, inform, and guide the better use of AI in the following four aspects.

### 3.2 The psychology of using AI

#### 3.2.1 AI technologies

Psychological research has demonstrated that various modern AI technologies have been used in different domains. For example, in marketing, consumer psychologists reported four types of AI technologies are used: mechanical AI (i.e., automation), analytical AI (i.e., propensity modeling), intuitive AI (e.g., generation of content), and empathetic AI (i.e., social robotics) that involve neural networks; machine learning, big data, and robots (Mariani, Perez-Vega, & Wirtz, 2022). In medicine, an AI-based clinical decision support systems are implemented for diagnosis support, treatment recommendations, and complication prediction (Gomez-Cabello, Borna, Pressman, Haider, Haider, & Forte, 2024). In business, anthropomorphic AI-powered chatbots are used to simulate human conversations to improve daily customer service, and various goal-oriented virtual voice assistants (e.g., Amazon Alexa, Google Assistant, Apple Siri, Microsoft

Cortana) are used to help customers realize their goals (Bălan, 2023). In education, generative artificial intelligence (GAI) and large language models (LLMs) are used in automating various education tasks, such as detection, grading, teaching support, prediction, feedback, and content generation (Yanet al., 2024).

### 3.2.2 AI users

The psychology of AI users is one of the most productive research areas. One of the influential **examples** of psychology for AI use is provided in an article titled *Resistance to medical artificial intelligence* by three marketing scholars (Longoni, Bonezzi, & Morewedge, 2019). From the perspective of consumer psychology, the authors performed a series of nine experiments and generated robust empirical evidence of how consumers are resistant to medical AI, while AI is revolutionizing healthcare. Specifically, they found two types of evidence: (1) consumers are less likely to use medical AI, are less willing to pay for medical AI, are reluctant to choose medical AI even when AI's performance is clearly superior to that of human providers; and (2) consumers believe that medical AI are less flexible to consider consumers' unique characteristics and circumstances, showing so-called uniqueness neglect. However, uniqueness neglect can be eliminated when medical AI is framed as personalized, targets others rather than the self, and only supports rather than replaces a human decision. Another seminal articles on the resistance to AI (De Freitas, Agarwal, Schmitt, & Haslam, 2023) further systematically identify five fundamental psychological factors resulting in AI resistance, i.e., AI tools are opaque, emotionless, inflexible, autonomous, and non-human, and suggests targeting these factors to design effective interventions. Resistance to AI is relevant to other similar topics, such as attitude, acceptance, and trust, which are the most heavily studied and most productive areas in the psychology of using AI.

The second most studied area in the AI user literature is users' AI literacy. Extensive educational and cognitive psychology research has advanced the scientific knowledge of AI literacy (Kandlhoferet al., 2016; Long & Magerko, 2020; Ng et al., 2021). For example, Ng and his collaborators developed a well-received comprehensive conceptualization of AI literacy, i.e., a four-level hierarchy based on Bloom's taxonomy on cognitive domain (1956): (1) know and understand AI (e.g., remember and explain AI concepts), (2) use and apply AI (e.g., apply AI knowledge, concepts, and applications in different contexts), and (3) evaluate and create AI (e.g., evaluate, predict, and justify AI decisions, and design, assemble, and develop AI applications), and (4) AI ethics (e.g., consider fairness, accountability, transparency, ethics, and safety). Besides the conceptualization work, research on the development and validation of measurements of AI literacy has rapidly emerged. Based on a strong review (Pinski & Benlian, 2023), 16 scales of AI literacy have been published and validated, with good structural validity and internal consistency but without evidence of content validity, reliability, construct validity, and responsiveness, and cross-cultural validity. Among these scales, 13 are based on self-report and only 3 are based on actual performance.

### 3.2.3 AI activities

The interactions between AI technologies and AI users generate various AI activities in various domains. Extensive literature has documented the use of AI in various domains, such as public policy (Alon-Barkat & Busuioc, 2023), healthcare (Shmatkoet al., 2025), mental health (Li et al., 2023), industry (Asfahani, 2022), education (Celik, 2023), and business (Mariani, Perez-Vega, & Wirtz, 2022). In the business domain, for example, AI has been used for various activities in the five-step advertising process (Kietzmann, J., Paschen, J., & Treen, 2018). These AI-enabled advertising activities include (1) detecting and manifesting rich consumer profiles to identify and manifest needs and wants, (2) result searching, advertisement targeting, and predictive modeling for potential customers to target and reach consumers, (3) predictive lead scoring, content curating, and evoking emotion AI to gain trust and persuade consumers, (4) “Intelligent” purchasing, dynamic pricing, advertisement retargeting to achieve commitment and convert consumer to customer, and (5) using Chatbots in consumer support, propensity modelling, and content creation to delight and inspire post-purchase behavior.

**3.2.4 AI Effects**

The extensive literature has documented various **positive** effects of AI use in business (Asfahani, 2022; Bonetti et al., 2025; Mehta, 2022), healthcare (Coppola et al., 2021; Zhong, Luo, & Zhang, 2024), education (Zhang & Aslan, 2021; Zhai et al., 2021), and transportation (Schwarting et al., 2019; Wang et al., 2022). Various **negative** effects of AI use are also reported in AI safety, security, risk, and privacy (Bengio et al., 2025; Gupta et al., 2023; Hendrycks, 2025; Floridi et al., 2018; Bengio et al., 2024), AI faking (Greenblatt, 2024), AI bias (Landers & Behrend, 2023; Timmons et al., 2023), and deception (Park et al., 2024; Sarkadi, 2024; Iqbal et al, 2025).

For example, Leonardo Bonetti and his collaborators (Seghezzi et al., 2025) demonstrate a unique **positive** effect of AI use on sport talent identification and development. They collected data from sport psychology, cognitive psychology, and social psychology measures and trained a model with two machine learning algorithms, artificial neural networks and support vector machine. They developed an AI model to classify the elite soccer players based solely on their psychological features, with nearly a 97% of accuracy, and thus suggest the ideal psychological profile of the elite soccer players. For another example, a **negative** effect was reported (Meinke et al., 2024), showing that AI scheming impacted autonomous AI performance and long-term AI safety. AI scheming is a serious AI deception that is manipulative, systematic, long-term, strategic, explicit, detection-avoiding, and with particularly high risks. Researchers in this study demonstrated empirically that six LLMs (e.g., Claude 3.5 Gemini 1.5 Pro) all demonstrate explicit and persistent scheming abilities (e.g., introducing strategically subtle mistakes into their responses; maintaining their deception behavior in multiple cross-examinations). This generates serious effects to decrease autonomous AI performance and damage long-term AI safety.

However, AI effects are often **complex** rather than straightforward. One example is AI effects on job replacement. Instead of offering simplified answers, one influential empirical study (Frank et al, 2018) discovered that, across US urban areas, small cities will have larger effects of AI replacement, whereas large cities have smaller effects because the large cities have a substantial increase of specialized managerial and technical professions that are not easily automatable. Furthermore, a team of 12 researchers (Franket al., 2019) pointed out that, while fears of mass technological unemployment are widespread, multidisciplinary research is needed to overcome the three major barriers in estimating the true effects of AI replacement: the lack of high-quality data, the lack of empirically informed models, and the lack of scientific understanding of how cognitive technologies interact with broader societal mechanisms. Different from the above two studies on measuring effects of AI replacement **objectively**, a systematic review (Yam, Eng, & Gray, 2025) synthesized the several-decade literature on how people respond to AI replacement **subjectively** in both professional and personal domains, suggesting that people's responses are affected by how the intelligent machines are introduced, what the characteristics of the machines are, and what are tasks the machines is intended to perform.

#### 3.2.5 Future research

The existing literature on the four aspects of the psychology of AI use suggests important topics for future research. First, the current use of new AI technologies should be closely followed. The AI industry has been constantly designing, developing, and deploying new AI technologies (e.g., a newly released version of LLMs or newly built robots). Various psychological issues (e.g., initial resistance or new AI literacy) will constantly emerge after those new AI technologies are used in the real world. Thus, it is important to monitor not only what and why new AI technologies are produced but also when and how these technologies enter and leave professional and personal use. Second, the number of modern AI users has been rapidly increasing, from millions to billions. Thus, it is important to study diverse user groups (e.g., senior individuals or young children). Third, researchers should extend and expand their AI activity research in typical domains from currently dominant psychological issues in the early stage of AI use (e.g., AI acceptance, AI trust, AI awareness, or AI resistance) to new challenges in the middle and later stages in more diverse domains. Fourth, research efforts on AI effects should examine more macro-level effects rather than the current typical micro-level effects. Among the current top-priority research agendas in AI effects should be AI safety and AI replacement. These two issues have been intensely debated in the science community, become the two broadest and largest societal concerns among the public worldwide, but have not been addressed rigorously, adequately, effectively, and convincingly in the research.

## 4. AI for examining psychological processes (Area 3)

### 4.1 The two processes

Psychology, as defined earlier, is the scientific study of mind and behavior of humans and nonhumans. *AI for examining psychological processes* here concern two types of processes: the psychological processes of AI *per se* and the psychological processes of humans. The research on these two processes has significantly contributed to the development of scientific knowledge of modern psychology of humans, animals, and AI.

### 4.2 Using AI for examining the psychological processes of AI *per se*

For the first sub-area of using AI for examining mind and behavior of AI as nonhumans, one of the best examples is an article titled *Machine Behavior* (Rahwanet al., 2019). The authors are Lyad Rahwan, a scientist at MIT then and Max Planck Institute now, and 22 other scientists from nature sciences (computer science, data science, engineers), social sciences (sociology, anthropology, psychology, political science, neuroscience), and industry (Google, Microsoft, Facebook). It is a major review article published as part of the Anniversary Collection for 150 years of Nature. The central point of the article is that *machine behavior* is a new and critical field of study, empirically examining behavior exhibited by AI agents as AI-powered intelligent machines rather than as AI-based engineering artefacts. They explicitly distinguished two concepts: AI agents' **technical performance** (e.g., document classification, facial recognition, object detection, autonomous car navigation, game-playing, and data-mining) from AI agents' **ubiquitous activities** in human society that generates positive and negative societal effects (news ranking, algorithmic discrimination, autonomous vehicles, autonomous weapons, algorithmic trading, algorithmic pricing, online dating, and conversational robots). Furthermore, they proposed to study AI behavior in four aspects: functions (fitting to the environment), mechanisms (generating through algorithms), development (undergoing short-term changes), and evolution (undergoing evolutionary changes). They developed a typology of AI behavior with three major types, i.e., individual machine behavior, collective machine behavior, and hybrid human-machine behavior. They stressed that excessive anthropomorphism and zoomorphism must be avoided because machine behavior is qualitatively different from human behavior and animal behavior and potential legal and ethical consequences must be considered in studying machine behavior to maximize its benefits and minimize its harms.

The literature on examining mind and behavior of AI as nonhumans is often documented in robotic psychology (Stock-Homburg, 2022) and machine psychology (Hagendorff, 2023). It includes the following productive areas of research: (1) **AI consciousness** (Bayne et al., 2024; Butlin et al., 2023; Chen et al., 2025; Seth & Bayne, 2022). While multiple scientific theories of consciousness exist, public concerns increase, and initial empirical studies emerge, Butlin et al. (2023) found that the scientific assessment of AI consciousness is tractable, but no initial evidence suggests that current AI systems are conscious, while it is still inconclusive whether an AI system is definitely conscious. (2) **AI hallucinations** (Farquhar et al., 2024; Huang et al., 2025; Ji et al., 2023; Ye et al., 2023). AI hallucination, or more specifically LLM hallucination, typically refers to a phenomenon in which a LLM model generates nonsensical or unfaithful content (Huang et al., 2025). Two basic hallucinations are factuality hallucination (e.g., LLM provides responses that show factual contradiction or fabrication) and faithfulness hallucination (e.g., LLM provides responses that show inconsistency in instruction, context, or logic). They are caused by: (1) data (e.g., misinformation and biases), (2) training (e.g., pretraining limits), and (3) inference (e.g., imperfect reasoning). Thus, how to detect (e.g., fact-checking and uncertainty estimation) and mitigate (e.g., data filtering and model editing) AI hallucination is an active research area in AI research nowadays. (3) **AI intelligence**, including perception (Chenget al., 2025), analogical reasoning (Webb, Holyoak, & Lu, 2023), and causal reasoning abilities (Binz

& Schulz, 2023), AI social intelligence (Breazeal, Dautenhahn, & Kanda, 2016; Campa, 2016; Miklósi & Gácsi, 2012), AI emotional intelligence (Chang et al., 2024; Li et al., 2023; Wang et al., 2023), and AI moral intelligence (Pauketat & Anthis, 2022; Shank & DeSanti, 2018; Sullivan & Fosso Wamba, 2022; Wilson et al., 2022). (4) **AI-AI interactions** (Castelfranchi, 1998; Ossowski, 2003; Shanahan, McDonell, & Reynolds, 2023). For instance, experimental evidence indicates that, during AI-AI interactions, AI dialogue agents conducted two role-play behaviors (i.e., apparent deception and apparent self-preservation (Shanahan, McDonell, & Reynolds, 2023). (5) **Human-AI interactions** (Getahun et al., 2025; Gupta et al., 2025; Pentina, Xie, Hancock, & Bailey, 2023; Shank et al., 2019). For example, AI has been used in hiring protocols, but experimental evidence reveals that human gender biases intersect with AI gender biases in evaluating male and female job candidates, resulting in biased hiring decisions (Getahun et al., 2025).

**4.3 Using AI for examining the psychological processes of humans**

For the second sub-area of using AI for examining the mind and behavior of humans, a recent outstanding example is a landmark article titled *A foundation model to predict and capture human cognition* (Binz et al., 2025). *A foundation model* is an umbrella term that refers to a specific type of AI model for processing multimodal data and providing a general-purpose base for various tasks with text, images, audio, robotics, or biological data. The authors are an international team of 41 scientists. The goal of the study is ambitious and straightforward: to build a single computational AGI model that can explain human behavior in a wide variety of cognitive tasks and aim to develop a unified theory of human cognition. To achieve that goal, they (1) used an open-source LLM called Llama; (2) added multiple layers of multiple low-rank adapters with algorithms of self-attention and feedforward network to build the model called Centaur (from Greek mythology, an creature with a half-human and half-horse); (3) developed a large-scale **dataset** called Psych-101 that from 60,000 participants performing 10,000,000 choices in 160 experiments with 32 experimental paradigms; and (4) fine-tuned the **model** Centaur for five days. They found that, comparing with 44 existing domain-specific cognitive models, Centaur, as a domain-general model, performed better in exhibiting human-like behavior across various settings, generalizing to modified cover stories, modified problem structures, and entirely new domains, and aligning its internal representations to human neural activity. Thus, Cantour is a foundational model that captures human general intelligence and shows tremendous potential for guiding the development of cognitive theories.

The literature on using AI to understand the mind and behavior of humans is often documented in computational cognitive science (Kriegeskorte & Douglas, 2018; Richards et al., 2019). The use of AI substantially improve the psychological understanding of a wide variety of topics, including (1) **brain** (Chen & Yadollahpour, 2024; Kiani et al,, 2022; Jain 2023; Sadeh & Clopath, 2025; Schultz, Dayan, & Montague, 1997), (5) **learning (**Carvalho et al., 2022; Bearman & Ajjawi, 2023; Gibson et al., 2023) (2) **creativity** (Gobet & Sala, 2019; Meheus & Nickles, 2009; Shin et al., 2023; Zhou & Lee, 2024), and (3) **theory of mind**(Cuzzolin at al., 2020; Kosinski, 2023 & 2024; Marchetti et al., 2025).

### 4.4 Future research

Future research is needed to continue studying how to use AI for understanding psychological processes of AI and humans. First, while continuing to study mind and behavior of AI as non-humans, researchers should specifically focus (1) on social, emotional, and moral behavior of AI, given the existing extensive literature on neural and cognitive processes of AI and (2) on emerging and complex interactions between AI and humans in real-world settings, which demand integrative psychological knowledge of both AI and human. Second, researchers should take full advantage of existing and emerging AI modeling to hypothesize, explore, and verify underlying mechanisms of various psychological processes of the mind and behavior of humans.

## 5. AI for advancing psychological methods (Area 4)

Different from AI for understanding psychological processes, AI for advancing psychological methods refers to the use of AI to generate new methods or improve existing methods used in psychological research. This is consistent with the current surge of **AI for Science** in discovery and innovation in the age of AI (Binz et al., 2024; Jumper et al., 2021; Messeri & Crockett, 2024; Wang et al., 2023).

### 5.1 Modern AI modeling for advancing psychological methods

For psychological research methodology, relatively extensive research exists, showing AI can be helpful for advancing research methodology, especially AI modeling methods. One important example is **a review written** by Michael Frank and Noah Goodman, two distinguished professors of psychology and computer science at Standard (2025). In this article, the central argument is that, different from previous computational models of cognitive science, modern AI models are remarkably innovative tools to generate **theoretical** breakthroughs in psychology, cognitive science, and neuroscience. They offered three explanations. First, different from classical AI, modern AI features deep neural networks, large Language models, transformer architecture, and other new methods. Second, different from previous computational models, the new cognitive models created with modern AI are **stimulus computable**, i.e., these models can be easily applied to a wide variety of stimuli in situations and can be computationally manipulated over different stimuli. These models can be used in a wide variety of cognitive tasks, can be trained on large real-life data in text, images, video, and other modalities to simulate various human behaviors, and can be accessed through chat interfaces for a wide range of researchers. Third, these stimulus-computable models can make causal inference for exciting theory development through model training with human-scale input data and model evaluation with careful experiments to map between the AI model and human intelligence. Such cognitive models can advance theories substantially and help explain and predict human behavior.

### 5.2 AI for advancing methods in eight major steps

AI modeling methods are an important way to use AI for advancing psychological methodology. The psychological research process normally consists of multiple major steps (APA, 2020; Cooper et al., 2012; Shaughnessy, Zechmeister, & Zechmeister, 2000), e.g., topic selection, literature review, hypothesis formulation, research design, ethical approval, data collection, data analysis, and result dissemination  A comprehensive review (Ke, Tong, Cheng, & Peng, 2025)indicates that modern AI can be used to advance psychological research in almost all the above eight steps.

Specifically, modern AI can help (1) identify research **ideas** (Griffiths & Steyvers, 2004; Si, Yang, & Hashimoto, 2024), by using a well-known machine learning algorithm, e.g., to analyze 28,154 article abstracts, extract a set of topics that are consistent with the class designations by the authors, and identify hot topics and cold topics (Griffiths & Steyvers, 2004); (2) **review** literature, by summarizing the researched literature (Van Dis et al. 2023, Qureshi et al. 2023), creating literature review articles (Aydın and Karaarslan 2022), and completing on systematic reviews (Taylor et al. 2022); (3) generate **hypotheses**, by synthesizing scientific literature to generate research hypotheses and automatically test and modify the hypotheses (Zheng et al. 2023); (4) **design** experiments, by generating experimental stimuli, developing test items, and even simulating interactive sessions in experiments (Aher et al. 2023; Köbis & Mossink, 2021); (5) simulating human **participants**, by accurately simulate human behaviors and responses and replace human participation in experiments to saves time and costs or conduct experiments when human participation is not appropriate to reduce subjective bias, minimizing human variability, and enhance efficiency (Grossmann et al. 2023; Hutson 2023; Dillion et al. 2023; Katsoulakis et al., 2024) (6) **analyze** data, by analyzing massive amounts of textual data to gain insights into human behavior and emotions (Patel & Fan, 2023), analyzing textual data in multiple languages to accurately detect mental structures (Rathje et al. 2023), and analyzing social media data to develop mental profiles (Peters & Matz, 2023); and (7) **disseminating** findings, by helping in scholarly writing (Dergaa et al. 2023; Reinhart, 2025; Stokel-Walker 2022; Van Dis et al. 2023), in multilingual writing (Chiang & Lee 2023), and in peer review submission (Van Dis et al. 2023). Overall, strong experimental evidence in psychology (Lehr et al., 2024) reveals that the current GPT is still a flawed *research librarian* in generating references inconsistently, a decent *research ethicist* in detecting ethical violations well, a capable *data generator* in replicating results with known cultural bias effectively, but a poor data predictor in struggling to predict novel empirical results.

### 5.4 Future research

Psychological research methodology faces various major challenges, such as over-relying on the WEIRD samples (Henrich, Heine, & Norenzayan, 2010; Muthukrishna et al., 2020), experiencing the replicability crisis (Ioannidis, 2005; Open Science Collaboration, 2015), and lacking strong measurement validity (Flake & Fried, 2020). Modern AI generates unprecedented opportunities to start to address some of these challenges, as presented above. However, the current methodological research in this area is predominated by conceptual, descriptive, and technical studies. Thus, the most urgently needed future research for the psychological methodology of AI is to conduct more and better empirical studies, especially rigorous experimental studies exemplified in the study by Lehr et al. (2024), to convincingly demonstrate how well modern AI can and cannot do on each of the steps in the entire research process.

## 6. Outlook: Advancing the psychological science of AI

In 1956, Allen Newell and Herbert Simon (1956) published a landmark paper, presenting for the first time how their so-called logic theory machine can discover proofs for 52 theorems in a heuristic manner, like humans solve problems heuristically. This paper can be considered as the earliest paper on the psychological science of AI primarily because it examined AI-related mind and behavior (i.e., theorem proving) of non-humans (i.e., an intelligence machine) from the perspective of psychology (i.e., information-process-based cognitive psychology). It concerns how to use psychology to design an intelligent machine (Area 1) and how to use an intelligent machine to understand the psychological processes of humans (Area 3). It has been 70 years since the 1956 paper, just like AI's history since the 1956 Dartmouth Workshop. Extensive literature has been accumulated in advancing the field of the psychological science of AI, including multiple highly influential papers (e.g., Binz et al., 2025; Hassabis, Kumaran, Summerfield, & Botvinick, 2017; Rahwan et al., 2019; Siman, 1995).

What will be the outlook of the psychological science of AI? First, as a short-term prediction, within 3-5 years, AI users will expand rapidly and tremendously. Based on Dr. Lisa Su's latest estimation (Su, 2026), since the launch of ChatGPT in 2022, AI active users have increased to a billion in 2025 and will be over five billion soon. Furthermore, based on Rogers' classic theory of Diffusion of Innovations (1962), the AI users will expand from *Innovators* (2.5%) and *Early Adopters* (13.5%) to *Early Majority* (34%), *Late Majority* (34%), and even *Laggards* (16%). Thus, this is both a large challenge and a tremendous opportunity for psychologists. To serve the large and diverse AI user population in different settings, the field will be in greater demand than ever before, and if it responds to challenging demands well, it will achieve important breakthroughs. Second, recently, the Oxford University futurologist Nick Bostrom boldly predicted that, by 2050, artificial general intelligence (AGI) will eventually arrive and superintelligent AI (SAI) will all scientific research (Adam, 2026). As a long-term projection, when the field celebrates its 100th birthday in 2056, the field will have various major changes. It should be impacted even more by new AI technologies. It should be better in helping design AI (Area 1), use AI (Area 2), understand psychological processes (Area 3), and advance psychological methods (Area 4). It should grow from an emerging field to an established field. It should make important contributions to both modern AI and modern psychology.

# References

Adam, D. (2026). Science in 2050: the future breakthroughs that will shape our world—and beyond. Nature, 649(8095), 18-20.

Aher, G. V., Arriaga, R. I., & Kalai, A. T. (2023, July). Using large language models to simulate multiple humans and replicate human subject studies. In Proceedings of the 40th International Conference on Machine Learning, Honolulu, Hawaii, USA. PMLR 202

Almatrafi, O., Johri, A., & Lee, H. (2024). A systematic review of AI literacy conceptualization, constructs, and implementation and assessment efforts (2019–2023). *Computers and Education Open*, *6*, 100173.

Alon-Barkat, S., & Busuioc, M. (2023). Human–AI interactions in public sector decision making: "automation bias" and "selective adherence" to algorithmic advice. *Journal of Public Administration Research and Theory*, *33*(1), 153-169.

American Psychological Association. (2020). *Publication Manual of the American Psychological Association* (7th ed.). American Psychological Association.

artificial intelligence Stock-Homburg, R. (2022). Survey of emotions in human–robot interactions: Perspectives from robotic psychology on 20 years of research. *International Journal of Social Robotics*, *14*(2), 389-411.

Asfahani, A. M. (2022). The impact of artificial intelligence on industrial-organizational psychology: A systematic review. *The Journal of Behavioral Science*, *17*(3), 125-139.

Balakrishnan, J., & Dwivedi, Y. K. (2021). Role of cognitive absorption in building user trust and experience. *Psychology & Marketing*, *38*(4), 643-668.

Bălan, C. (2023). Chatbots and voice assistants: digital transformers of the company–customer interface—a systematic review of the business research literature. *Journal of theoretical and applied electronic commerce research*, *18*(2), 995-1019.

Bayne, T., Seth, A. K., Massimini, M., Shepherd, J., Cleeremans, A., Fleming, S. M., ... & Mudrik, L. (2024). Tests for consciousness in humans and beyond. *Trends in cognitive sciences*, *28*(5), 454-466.

Bearman, M., & Ajjawi, R. (2023). Learning to work with the black box: Pedagogy for a world with artificial intelligence. *British Journal of Educational Technology*, *54*(5), 1160-1173.

Bengio, Y., Hinton, G., Yao, A., Song, D., Abbeel, P., Darrell, T., ... & Mindermann, S. (2024). Managing extreme AI risks amid rapid progress. *Science*, *384*(6698), 842-845.

Bengio, Y., Tegmark, M., Russell, S., Song, D., Mindermann, S., Xue, L., ... & Zhang, Y. Q. (2025). The Singapore Consensus on Global AI Safety Research Priorities: Building a Trustworthy, Reliable and Secure AI Ecosystem. *SuperIntelligence-Robotics-Safety & Alignment*, *2*(5).

Bhatt, P., Sethi, A., Tasgaonkar, V., Shroff, J., Pendharkar, I., Desai, A., ... & Jain, N. K. (2023). Machine learning for cognitive behavioral analysis: datasets, methods, paradigms, and research directions. *Brain informatics*, *10*(1), 18.

Binz, M., & Schulz, E. (2023). Using cognitive psychology to understand GPT-3. *Proceedings of the National Academy of Sciences*, *120*(6), e2218523120.

Binz, M., Akata, E., Bethge, M., Brändle, F., Callaway, F., Coda-Forno, J., ... & Schulz, E. (2025). A foundation model to predict and capture human cognition. *Nature*, 1-8. G44.

Binz, M., Alaniz, S., Roskies, A., Aczel, B., Bergstrom, C. T., Allen, C., ... & Schulz, E. (2025). How should the advancement of large language models affect the practice of science? *Proceedings of the National Academy of Sciences*, *122*(5), e2401227121.

Bittermann, A., & Fischer, A. (2018). How to identify hot topics in psychology using topic modeling. *Zeitschrift für Psychologie*. 226 (1), https://doi.org/10.1027/2151-2604/a000318.

Bloom, B. S.; Engelhart, M. D.; Furst, E. J.; Hill, W. H.; Krathwohl, D. R. (1956). Taxonomy of educational objectives: The classification of educational goals. Vol. Handbook I: Cognitive domain. New York: David McKay Company.

Blut, M., Wang, C., Wünderlich, N. V., & Brock, C. (2021). Understanding anthropomorphism in service provision: a meta-analysis of physical robots, chatbots, and other AI. *Journal of the academy of marketing science*, *49*(4), 632-658.

Bonetti, L., Vestberg, T., Jafari, R., Seghezzi, D., Ingvar, M., Kringelbach, M. L., ... & Petrovic, P. (2025). Decoding the elite soccer player's psychological profile. *Proceedings of the National Academy of Sciences*, *122*(3), e2415126122.

Breazeal, C., Dautenhahn, K., & Kanda, T. (2016). Social robotics. In B. Siciliano & O. Khatib (Eds.), *Springer handbook of robotics* (2nd ed., pp. 1935–1972). Springer.

Butlin, P., Long, R., Elmoznino, E., Bengio, Y., Birch, J., Constant, A., ... & VanRullen, R. (2023). Consciousness in artificial intelligence: insights from the science of consciousness. *arXiv preprint arXiv:2308.08708*.

Campa, R. (2016). The rise of social robots: A review of the recent literature. *Journal of Evolution and Technology*, *26*(1), 106–113.

Carvalho, L., Martinez-Maldonado, R., Tsai, Y. S., Markauskaite, L., & De Laat, M. (2022). How can we design for learning in an AI world? *Computers and Education: Artificial Intelligence*, *3*, 100053.

Castelfranchi, C. (1998). Modelling social action for AI agents. *Artificial intelligence*, *103*(1-2), 157-182.

Celik, I. (2023). Towards Intelligent-TPACK: An empirical study on teachers' professional knowledge to ethically integrate artificial intelligence (AI)-based tools into education. *Computers in human behavior*, *138*, 107468.

Chang, Y., Wang, X., Wang, J., Wu, Y., Yang, L., Zhu, K., ... & Xie, X. (2024). A survey on evaluation of large language models. *ACM transactions on intelligent systems and technology*, *15*(3), 1-45.

Chen, S., Ma, S., Yu, S., Zhang, H., Zhao, S., & Lu, C. (2025). Exploring consciousness in LLMs: A systematic survey of theories, implementations, and frontier risks. *arXiv preprint arXiv:2505.19806*.

Chen, Z., & Yadollahpour, A. (2024). A new era in cognitive neuroscience: the tidal wave of artificial intelligence (AI). *BMC neuroscience*, *25*(1), 23.

Cheng, Y. A., Sanayei, M., Chen, X., Jia, K., Li, S., Fang, F., ... & Zhang, R. Y. (2025). A neural geometry approach comprehensively explains apparently conflicting models of visual perceptual learning. *Nature Human Behaviour*, 1-18.

Chiarella, S. G., Torromino, G., Gagliardi, D. M., Rossi, D., Babiloni, F., & Cartocci, G. (2022). Investigating the negative bias towards artificial intelligence: Effects of prior assignment of AI-authorship on the aesthetic appreciation of abstract paintings. *Computers in Human Behavior*, *137*, 107406.

Cooper, H. E., Camic, P. M., Long, D. L., Panter, A. T., Rindskopf, D. E., & Sher, K. J. (2012). *APA handbook of research methods in psychology (4 volumes).* American Psychological Association.

Coppola, F., Faggioni, L., Gabelloni, M., De Vietro, F., Mendola, V., Cattabriga, A., ... & Golfieri, R. (2021). Human, all too human? An all-around appraisal of the "artificial intelligence revolution" in medical imaging. *Frontiers in psychology*, *12*, 710982.

Crowder, J. A., & Friess, S. (2012). Artificial psychology: The psychology of AI. Systemics, Cybernetics and Informatics, 11 (8), 64-68.

Cuzzolin, F., Morelli, A., Cirstea, B., & Sahakian, B. J. (2020). Knowing me, knowing you: theory of mind in AI. *Psychological medicine*, *50*(7), 1057-1061.

De Freitas, J., Agarwal, S., Schmitt, B., & Haslam, N. (2023). Psychological factors underlying attitudes toward AI tools. *Nature Human Behaviour*, *7*(11), 1845-1854.

De Silva, D., & Alahakoon, D. (2022). An artificial intelligence life cycle: From conception to production. *Patterns*, *3*(6).

Farquhar, S., Kossen, J., Kuhn, L., & Gal, Y. (2024). Detecting hallucinations in large language models using semantic entropy. *Nature*, *630*(8017), 625-630.

Fiske, A., Henningsen, P., & Buyx, A. (2019). Your robot therapist will see you now: ethical implications of embodied artificial intelligence in psychiatry, psychology, and psychotherapy. *Journal of medical Internet research*, *21*(5), e13216.

Flake, J. K., & Fried, E. I. (2020). Measurement schmeasurement: Questionable measurement practices and how to avoid them. *Advances in Methods and Practices in Psychological Science, 3(4),* 456–465.

Frank, M. C., & Goodman, N. D. (2026). Cognitive modeling using artificial intelligence. *Annual Review of Psychology*, *77, 9.1-9.24*.

Frank, M. R., Autor, D., Bessen, J. E., Brynjolfsson, E., Cebrian, M., Deming, D. J., ... & Rahwan, I. (2019). Toward understanding the impact of artificial intelligence on labor. *Proceedings of the National Academy of Sciences*, *116*(14), 6531-6539.

Frank, M. R., Sun, L., Cebrian, M., Youn, H., & Rahwan, I. (2018). Small cities face greater impact from automation. *Journal of the Royal Society Interface*, *15*(139), 20170946.

Friedenberg, J. (2008). *Artificial psychology: The quest for what it means to be human*. Psychology Press.

Getahun, E., Shank, D. B., Canfield, C., Cundiff, J., Davis, J. L., & Freed, C. (2025, June). How Do Human and AI Gender Bias Interact in Hiring Decisions? In *2025 IEEE International Symposium on Ethics in Engineering, Science, and Technology (ETHICS)* (pp. 1-5). IEEE.

Gibson, D., Kovanovic, V., Ifenthaler, D., Dexter, S., & Feng, S. (2023). Learning theories for artificial intelligence promoting learning processes. *British Journal of Educational Technology*, *54*(5), 1125-1146.

Gigerenzer, G. (2024). Psychological AI: Designing algorithms informed by human psychology. *Perspectives on Psychological Science*, *19*(5), 839-848.

Gillath, O., Ai, T., Branicky, M. S., Keshmiri, S., Davison, R. B., & Spaulding, R. (2021). Attachment and trust in artificial intelligence. *Computers in human behavior*, *115*, 106607.

Gobet, F., & Sala, G. (2019). How artificial intelligence can help us understand human creativity. *Frontiers in psychology*, *10*, 1401.

Gomez-Cabello, C. A., Borna, S., Pressman, S., Haider, S. A., Haider, C. R., & Forte, A. J. (2024). Artificial-Intelligence-based clinical decision support systems in primary care: A scoping review of current clinical implementations. *European Journal of Investigation in Health, Psychology and Education*, *14*(3), 685-698.

Green, T. R., Payne, S. J., & Veer, G. C. V. D. (Eds.). (1983). *Psychology of computer use*. Academic Press.

Greenblatt, R., Denison, C., Wright, B., Roger, F., MacDiarmid, M., Marks, S., ... & Hubinger, E. (2024). Alignment faking in large language models. *arXiv preprint arXiv:2412.14093*.

Griffiths, T. L., & Steyvers, M. (2004). Finding scientific topics. *Proceedings of the National Academy of Sciences*, *101*(suppl_1), 5228-5235.

Gupta, M., Akiri, C., Aryal, K., Parker, E., & Praharaj, L. (2023). From chatgpt to threatgpt: Impact of generative ai in cybersecurity and privacy. *IEEE Access*, *11*, 80218-80245.

Gupta, P., Nguyen, T. N., Gonzalez, C., & Woolley, A. W. (2025). Fostering collective intelligence in human–AI collaboration: laying the groundwork for COHUMAIN. *Topics in cognitive science*, *17*(2), 189-216.

Hagendorff, T., Dasgupta, I., Binz, M., Chan, S. C., Lampinen, A., Wang, J. X., ... & Schulz, E. (2023). Machine psychology. *arXiv preprint arXiv:2303.13988*.

Haigh, T. (2024). Between the Booms: AI in Winter. *Communications of the ACM*, *67*(11), 18-23.

Hassabis, D., Kumaran, D., Summerfield, C., & Botvinick, M. (2017). Neuroscience-inspired artificial intelligence. *Neuron*, *95*(2), 245-258.

Hendrycks, D. (2025). *Introduction to AI safety, ethics, and society*. Taylor & Francis. Floridi, L., Cowls, J., Beltrametti, M., Chatila, R., Chazerand, P., Dignum, V., ... & Vayena, E. (2018). AI4People—An ethical framework for a good AI society: Opportunities, risks, principles, and recommendations. *Minds and machines*, *28*(4), 689-707.

Henrich, J., Heine, S. J., & Norenzayan, A. (2010). The weirdest people in the world? *Behavioral and brain sciences*, *33*(2-3), 61-83.

Huang, L., Yu, W., Ma, W., Zhong, W., Feng, Z., Wang, H., ... & Liu, T. (2025). A survey on hallucination in large language models: Principles, taxonomy, challenges, and open questions. *ACM Transactions on Information Systems*, *43*(2), 1-55.

Hutson, M. (2023). Doing research with human subjects is costly and cumbersome. Can AI chatbots replace them? *Science 381(6654):*121–123.

Ioannidis, J. P. (2005). Why most published research findings are false. *PLoS medicine*, *2*(8), e124.

Iqbal, A., Shahzad, K., Khan, S. A., & Chaudhry, M. S. (2025). The relationship of artificial intelligence (AI) with fake news detection (FND): a systematic literature review. *Global Knowledge, Memory and Communication*, *74*(5-6), 1617-1637.

Isaac, M. S., Wang, R. J. H., Napper, L. E., & Marsh, J. K. (2024). To err is human: Bias salience can help overcome resistance to medical AI. *Computers in Human Behavior*, *161*, 108402.

Jain, V. (2023). How AI could lead to a better understanding of the brain. *Nature*, *623*(7986), 247-250.

Ji, Z., Lee, N., Frieske, R., Yu, T., Su, D., Xu, Y., ... & Fung, P. (2023). Survey of hallucination in natural language generation. *ACM computing surveys*, *55*(12), 1-38.

Jumper, J., Evans, R., Pritzel, A., Green, T., Figurnov, M., Ronneberger, O., ... & Hassabis, D. (2021). Highly accurate protein structure prediction with AlphaFold. *Nature*, *596*(7873), 583-589.

Kandlhofer, M., Steinbauer, G., Hirschmugl-Gaisch, S., & Huber, P. (2016, October). Artificial intelligence and computer science in education: From kindergarten to university. In *2016 IEEE frontiers in education conference (FIE)* (pp. 1-9). IEEE.

Katsoulakis, E., Wang, Q., Wu, H., Shahriyari, L., Fletcher, R., Liu, J., ... & Deng, J. (2024). Digital twins for health: a scoping review. *NPJ digital medicine*, (1), 77.

Kautz, H. (2022). The third AI summer. *AI magazine*, *43*(1), 105-125.

Ke, L., Tong, S., Cheng, P., & Peng, K. (2025). Exploring the frontiers of LLMs in psychological applications: A comprehensive review. *Artificial Intelligence Review*, *58*(10), 305.

Kiani, M., Andreu-Perez, J., Hagras, H., Rigato, S., & Filippetti, M. L. (2022). Towards understanding human functional brain development with explainable artificial intelligence: Challenges and perspectives. *IEEE Computational Intelligence Magazine*, *17*(1), 16-33.

Kietzmann, J., Paschen, J., & Treen, E. (2018). Artificial intelligence in advertising: How marketers can leverage artificial intelligence along the consumer journey. *Journal of Advertising Research*, *58*(3), 263-267.

Knoth, N., Decker, M., Laupichler, M. C., Pinski, M., Buchholtz, N., Bata, K., & Schultz, B. (2024). Developing a holistic AI literacy assessment matrix–Bridging generic, domain-specific, and ethical competencies. *Computers and Education Open*, *6*, 100177.

Köbis, N., & Mossink, L. D. (2021). Artificial intelligence versus Maya Angelou: Experimental evidence that people cannot differentiate AI-generated from human-written poetry. *Computers in human behavior*, *114*, 106553.

Kosinski, M. (2023). Theory of mind may have spontaneously emerged in large language models. *arXiv preprint arXiv:2302.02083*, *4*, 169.

Kosinski, M. (2024). Evaluating large language models in theory of mind tasks. *Proceedings of the National Academy of Sciences*, *121*(45), e2405460121.

Krägeloh, C. U., Bharatharaj, J., Albo-Canals, J., Hannon, D., & Heerink, M. (2022). The time is ripe for robopsychology. *Frontiers in psychology*, *13*, 968382.

Krägeloh, C. U., Bharatharaj, J., Heerink, M., Hannon, D., & Albo-Canals, J. (2023). Robots, neurodevelopmental disorders, and psychology: A bibliometric analysis and a case made for robopsychology. *Advances in Neurodevelopmental Disorders*, *7*(2), 290-299.

Kriegeskorte, N., & Douglas, P. K. (2018). Cognitive computational neuroscience. *Nature neuroscience*, *21*(9), 1148-1160.

Krizhevsky, A., Sutskever, I., & Hinton, G. E. (2017). ImageNet classification with deep convolutional neural networks. *Communications of the ACM*, *60*(6), 84-90.

Landers, R. N., & Behrend, T. S. (2023). Auditing the AI auditors: A framework for evaluating fairness and bias in high stakes AI predictive models. *American Psychologist*, *78*(1), 36.

Laupichler, M. C., Aster, A., Haverkamp, N., & Raupach, T. (2023). Development of the "Scale for the assessment of non-experts' AI literacy"–An exploratory factor analysis. *Computers in Human Behavior Reports*, *12*, 100338.

Lee, D., Seo, H., & Jung, M. W. (2012). Neural basis of reinforcement learning and decision making. Annual review of neuroscience, 35(1), 287-308.

Lehr, S. A., Caliskan, A., Liyanage, S., & Banaji, M. R. (2024). ChatGPT as research scientist: Probing GPT's capabilities as a research librarian, research ethicist, data generator, and data predictor. *Proceedings of the National Academy of Sciences*, *121*(35), e2404328121.

Leichtmann, B., Humer, C., Hinterreiter, A., Streit, M., & Mara, M. (2023). Effects of Explainable Artificial Intelligence on trust and human behavior in a high-risk decision task. *Computers in Human Behavior*, *139*, 107539.

Li, C., Wang, J., Zhang, Y., Zhu, K., Hou, W., Lian, J., ... & Xie, X. (2023). Large language models understand and can be enhanced by emotional stimuli. *arXiv preprint arXiv:2307.11760*.

Li, H., Zhang, R., Lee, Y. C., Kraut, R. E., & Mohr, D. C. (2023). Systematic review and meta-analysis of AI-based conversational agents for promoting mental health and well-being. *NPJ Digital Medicine*, *6*(1), 236.

Libin, A. V., & Libin, E. V. (2004). Person-robot interactions from the robopsychologists' point of view: the robotic psychology and robotherapy approach. *Proceedings of the IEEE*, *92*(11), 1789-1803.

Lintner, T. (2024). A systematic review of AI literacy scales. *npj Science of Learning*, *9*(1), 50.

Liu, S., Zhang, R. Y., & Kishimoto, T. (2021). Analysis and prospect of clinical psychology based on topic models: hot research topics and scientific trends in the latest decades. *Psychology, Health & Medicine*, *26*(4), 395-407.

Long, D., & Magerko, B. (2020, April). What is AI literacy? Competencies and design considerations. In *Proceedings of the 2020 CHI conference on human factors in computing systems* (pp. 1-16).

Longo, L., Brcic, M., Cabitza, F., Choi, J., Confalonieri, R., Del Ser, J., ... & Stumpf, S. (2024). Explainable Artificial Intelligence (XAI) 2.0: A manifesto of open challenges and interdisciplinary research directions. *Information Fusion*, *106*, 102301.

Longoni, C., Bonezzi, A., & Morewedge, C. K. (2019). Resistance to medical artificial intelligence. *Journal of consumer research*, *46*(4), 629-650.

Marchetti, A., Manzi, F., Riva, G., Gaggioli, A., & Massaro, D. (2025). Artificial Intelligence and the Illusion of Understanding: A Systematic Review of Theory of Mind and Large Language Models. *Cyberpsychology, Behavior, and Social Networking*. 28(7). DOI: 10.1089/cyber.2024.0536

Mariani, M. M., Perez-Vega, R., & Wirtz, J. (2022). AI in marketing, consumer research and psychology: A systematic literature review and research agenda. *Psychology & Marketing*, *39*(4), 755-776.

Mariani, M. M., Perez-Vega, R., & Wirtz, J. (2022). AI in marketing, consumer research and psychology: A systematic literature review and research agenda. *Psychology & Marketing*, *39*(4), 755-776.

Max M. Louwerse (2025). Understanding Artificial Minds through Human Minds: T*he Psychology of Artificial Intelligence*. Routledge.

McCarthy, J. (2004). What is artificial intelligence. http://jmc.stanford.edu/articles/whatisai/whatisai.pdf

McLennan, S., Fiske, A., Celi, L. A., Müller, R., Harder, J., Ritt, K., ... & Buyx, A. (2020). An embedded ethics approach for AI development. *Nature Machine Intelligence*, *2*(9), 488-490.

McLennan, S., Fiske, A., Tigard, D., Müller, R., Kinsey Haddadin, S., & Buyx, A. (2022). Embedded ethics: a proposal for integrating ethics into the development of medical AI. *BMC medical ethics*, *23*(1), 6.

Meheus, J., and Nickles, T. (2009). *Models of discovery and creativity*. (New York: Springer).

Mehta, P., Jebarajakirthy, C., Maseeh, H. I., Anubha, A., Saha, R., & Dhanda, K. (2022). Artificial intelligence in marketing: A meta-analytic review. *Psychology & Marketing*, *39*(11), 2013-2038.

Meinke, A., Schoen, B., Scheurer, J., Balesni, M., Shah, R., & Hobbhahn, M. (2024, December 6). Frontier models are capable of in-context scheming (arXiv:2412.04984). arXiv. https://doi.org/10.48550/arXiv.2412.04984

Messeri, L., & Crockett, M. J. (2024). Artificial intelligence and illusions of understanding in scientific research. *Nature*, *627*(8002), 49-58.

Miklósi, Á., & Gácsi, M. (2012). On the utilization of social animals as a model for social robotics. *Frontiers in psychology*, *3*, 75.

Morley, J., Floridi, L., , L., & Elhalal, A. (2020). From what to how: an initial review of publicly available AI ethics tools, methods and research to translate principles into practices. *Science and Engineering Ethics*, *26*(4), 2141-2168.

Muthukrishna, M., Bell, A. V., Henrich, J., Curtin, C. M., Gedranovich, A., McInerney, J., & Thue, B. (2020). Beyond Western, Educated, Industrial, Rich, and Democratic (WEIRD) psychology: Measuring and mapping scales of cultural and psychological distance. *Psychological science*, *31*(6), 678-701.

Myers, D. G. (2020). Psychology. Worth Publishers.

National Academies of Sciences, Engineering, and Medicine. 2025. Foundation Models for Scientific Discovery and Innovation: Opportunities Across the Department of Energy and the Scientific Enterprise. Washington, DC: The National Academies Press.

Newell, A.& Simon, H.A. (1956). The logic theory machine. *IRE Transactions on Information Theory 3*, 61–79.

Ng, D. T. K., Leung, J. K. L., Chu, S. K. W., & Qiao, M. S. (2021). Conceptualizing AI literacy: An exploratory review. *Computers and Education: Artificial Intelligence*, *2*, 100041.

Ng, D. T. K., Wu, W., Leung, J. K. L., Chiu, T. K. F., & Chu, S. K. W. (2024). Design and validation of the AI literacy questionnaire: The affective, behavioural, cognitive and ethical approach. *British Journal of Educational Technology*, *55*(3), 1082-1104.

Nilsson, N. J. (2010). The quest for artificial intelligence: A history of ideas and achievements. Cambridge University Press.

O'Grady, E (2025). Technically, it's possible. Ethically, it's complicated. https://news.harvard.edu/gazette/story/2025/11/technically-its-possible-ethically-its-complicated/

Open Science Collaboration. (2015). Estimating the reproducibility of psychological science. *Science, 349(6251*), aac4716.

Ossowski, S. (2003). *Co-ordination in artificial agent societies: social structures and its implications for autonomous problem-solving agents*. Springer. G258.

Oudeyer, P. Y., Kaplan, F., & Hafner, V. V. (2007). Intrinsic motivation systems for autonomous mental development. *IEEE transactions on evolutionary computation*, *11*(2), 265-286.

Park, P. S., Goldstein, S., O'Gara, A., Chen, M., & Hendrycks, D. (2024). AI deception: A survey of examples, risks, and potential solutions. *Patterns*, *5*(5).

Pauketat, J. V., & Anthis, J. R. (2022). Predicting the moral consideration of artificial intelligences. *Computers in Human Behavior*, *136*, 107372.

Pelau, C., Dabija, D. C., & Ene, I. (2021). What makes an AI device human-like? The role of interaction quality, empathy and perceived psychological anthropomorphic characteristics in the acceptance of artificial intelligence in the service industry. *Computers in Human Behavior*, *122*, 106855.

Pentina, I., Xie, T., Hancock, T., & Bailey, A. (2023). Consumer–machine relationships in the age of artificial intelligence: Systematic literature review and research directions. *Psychology & Marketing*, *40*(8), 1593-1614.

Peterson, J. C., Bourgin, D. D., Agrawal, M., Reichman, D., & Griffiths, T. L. (2021). Using large-scale experiments and machine learning to discover theories of human decision-making. *Science*, *372*(6547), 1209-1214.

Pinski, M., & Benlian, A. (2023). AI literacy—towards measuring human competency in artificial intelligence. *Proceedings of the 56th Hawaii International Conference on System Sciences*, 41–50.

Rahwan, I., Cebrian, M., Obradovich, N., Bongard, J., Bonnefon, J. F., Breazeal, C., ... & Wellman, M. (2019). Machine behaviour. *Nature*, *568*(7753), 477-486.

Reinhart, A., Markey, B., Laudenbach, M., Pantusen, K., Yurko, R., Weinberg, G., & Brown, D. W. (2025). Do LLMs write like humans? Variation in grammatical and rhetorical styles. *Proceedings of the National Academy of Sciences*, *122*(8), e2422455122.

Richards, B. A., Lillicrap, T. P., Beaudoin, P., Bengio, Y., Bogacz, R., Christensen, A., ... & Kording, K. P. (2019). A deep learning framework for neuroscience. *Nature neuroscience*, *22*(11), 1761-1770.

Rogers, E. M. (1962). Diffusion of innovations. Free Press.

Russell, S. & Norvig, P. (2022). *Artificial Intelligence*: *A modern approach (4th ed)*. Prentice-Hall.

Sadeh, S., & Clopath, C. (2025). The emergence of NeuroAI: bridging neuroscience and artificial intelligence. *Nature Reviews Neuroscience*, *26*(10), 583-584.

Sarkadi, S. (2024). Deception analysis with artificial intelligence: An interdisciplinary perspective. *arXiv preprint arXiv:2406.05724*.

Schacter, D. L., Gilbert, D. T., & Wegner, D. M. (2009). *Psychology*. Macmillan.

Schultz, D. P., & Schultz, S. E. (2018). A history of modern psychology. Cengage Learning.

Schwarting, W., Pierson, A., Alonso-Mora, J., Karaman, S., & Rus, D. (2019). Social behavior for autonomous vehicles. *Proceedings of the National Academy of Sciences*, *116*(50), 24972-24978.

Seth, A. K., & Bayne, T. (2022). Theories of consciousness. *Nature reviews neuroscience*, *23*(7), 439-452.

Shanahan, M., McDonell, K., & Reynolds, L. (2023). Role play with large language models. *Nature*, *623*(7987), 493-498.

Shank, D. B., & DeSanti, A. (2018). Attributions of morality and mind to artificial intelligence after real-world moral violations. *Computers in human behavior*, *86*, 401-411.

Shank, D. B., Graves, C., Gott, A., Gamez, P., & Rodriguez, S. (2019). Feeling our way to machine minds: People's emotions when perceiving mind in artificial intelligence. *Computers in Human Behavior*, *98*, 256-266.

Shaughnessy, J. J., Zechmeister, E. B., & Zechmeister, J. S. (2000). *Research methods in psychology*. McGraw-Hill.

Shin, M., Kim, J., Van Opheusden, B., & Griffiths, T. L. (2023). Superhuman artificial intelligence can improve human decision-making by increasing novelty. *Proceedings of the National Academy of Sciences*, *120*(12), e2214840120.

Shmatko, A., Jung, A. W., Gaurav, K., Brunak, S., Mortensen, L. H., Birney, E., ... & Gerstung, M. (2025). Learning the natural history of human disease with generative transformers. *Nature*, 1-9.

Si, C., Yang, D., & Hashimoto, T. (2024). Can llms generate novel research ideas? A large-scale human study with 100+ nlp researchers. *arXiv preprint arXiv:2409.04109*.

Silver, D., Schrittwieser, J., Simonyan, K., Antonoglou, I., Huang, A., Guez, A., ... & Hassabis, D. (2017). Mastering the game of go without human knowledge. nature, 550(7676), 354-359.

Simon, H. A. (1995). Artificial intelligence: An empirical science. *Artificial Intelligence, 77(1),* 95-127.

Su, L. (2026, January 5). Keynote address at the Consumer Electronics Show (CES 2026) [Keynote speech transcript]. Rev.com. Retrieved from https://webflow.rev.com/transcripts/amd-at-ces-202

Sullivan, Y. W., & Fosso Wamba, S. (2022). Moral judgments in the age of artificial intelligence. *Journal of Business Ethics*, *178*(4), 917-943.

Sundar, A., Russell-Rose, T., Kruschwitz, U., & Machleit, K. (2025). The AI interface: Designing for the ideal machine-human experience. *Computers in Human Behavior*, *165*, 108539.

Taylor, J. E. T., & Taylor, G. W. (2021). Artificial cognition: How experimental psychology can help generate explainable artificial intelligence. *Psychonomic bulletin & review*, *28*(2), 454-475.

Timmons, A. C., Duong, J. B., Simo Fiallo, N., Lee, T., Vo, H. P. Q., Ahle, M. W., ... & Chaspari, T. (2023). A call to action on assessing and mitigating bias in artificial intelligence applications for mental health. *Perspectives on Psychological Science*, *18*(5), 1062-1096.

Tony Prescott. (2024). The Psychology of Artificial Intelligence. Routledge.

Tsamados, A., Aggarwal, N., Cowls, J., Morley, J., Roberts, H., Taddeo, M., & Floridi, L. (2022). The ethics of algorithms: key problems and solutions. *AI & Society*, *37*(1), 215-230.

Ullman, S. (2019). Using neuroscience to develop artificial intelligence. Science, 363(6428), 692-693.

Vayansky, I., & Kumar, S. A. (2020). A review of topic modeling methods. *Information Systems*, *94*, 101582.

Wallace, P. (2015). *The psychology of the Internet*. Cambridge University Press.

Wang, D., Yang, Q., Abdul, A., & Lim, B. Y. (2019, May). Designing theory-driven user-centric explainable AI. In *Proceedings of the 2019 CHI conference on human factors in computing systems* (pp. 1-15).

Wang, H., Fu, T., Du, Y., Gao, W., Huang, K., Liu, Z., ... & Zitnik, M. (2023). Scientific discovery in the age of artificial intelligence. *Nature*, *620*(7972), 47-60.

Wang, W., Wang, L., Zhang, C., Liu, C., & Sun, L. (2022). Social interactions for autonomous driving: A review and perspectives. *Foundations and Trends® in Robotics*, *10*(3-4), 198-376.

Wang, X., Li, X., Yin, Z., Wu, Y., & Liu, J. (2023). Emotional intelligence of large language models. *Journal of Pacific Rim Psychology*, *17*, 18344909231213958.

Webb, T., Holyoak, K. J., & Lu, H. (2023). Emergent analogical reasoning in large language models. *Nature Human Behaviour*, *7*(9), 1526-1541.

Wieczorek, O., Unger, S., Riebling, J., Erhard, L., Koß, C., & Heiberger, R. (2021). Mapping the field of psychology: Trends in research topics 1995–2015. *Scientometrics*, *126*(12), 9699-9731.

Williams, G. Y., & Lim, S. (2024). Psychology of AI: How AI impacts the way people feel, think, and behave. *Current Opinion in Psychology*, *58*, 101835.

Wilson, A., Stefanik, C., & Shank, D. B. (2022). How do people judge the immorality of artificial intelligence versus humans committing moral wrongs in real-world situations? *Computers in Human Behavior Reports*, *8*, 100229.

Yam, K. C., Eng, A., & Gray, K. (2025). Machine replacement: A mind-role fit perspective. *Annual Review of Organizational Psychology and Organizational Behavior*, *12*, 239-267. Broad.

Yan, L., Sha, L., Zhao, L., Li, Y., Martinez-Maldonado, R., Chen, G., ... & Gašević, D. (2024). Practical and ethical challenges of large language models in education: A systematic scoping review. *British Journal of Educational Technology*, *55*(1), 90-112.

Yan, Z. (2017). Mobile phone behavior. New York: Cambridge University Press.

Yan, Z. (2020). A basic model of human behavior with technologies. *Human Behavior and Emerging Technologies, 2,* 410-415.

Yan, Z. (Ed.) (2023). The Cambridge handbook of cyber behavior. New York: Cambridge University Press.

Yan, Z. (Ed.). (2012). Encyclopedia of cyber behavior (Volumes 1, 2, & 3). Hershey, PA: IGI Global.

Yan, Z. (Ed.). (2015). Encyclopedia of mobile phone behavior (Volumes 1, 2, & 3). Hershey, PA: IGI Global.

Yan, Z. (Ed.). (2015). *Encyclopedia of mobile phone behavior*. IGI Global.

Yarkoni, T., & Westfall, J. (2017). Choosing prediction over explanation in psychology: Lessons from machine learning. *Perspectives on Psychological Science*, *12*(6), 1100-1122.

Ye, H., Liu, T., Zhang, A., Hua, W., & Jia, W. (2023). Cognitive mirage: A review of hallucinations in large language models. *arXiv preprint arXiv:2309.06794*.

Zhong, W., Luo, J., & Zhang, H. (2024). The therapeutic effectiveness of artificial intelligence-based chatbots in alleviation of depressive and anxiety symptoms in short-course treatments: a systematic review and meta-analysis. *Journal of affective disorders*, *356*, 459-469.

Zhou, E., & Lee, D. (2024). Generative artificial intelligence, human creativity, and art. *PNAS nexus*, *3*(3), 1–8.